\documentclass[aps,preprint,showpacs]{revtex4}

\usepackage{graphicx,epsfig}
\usepackage{amssymb}
\begin{document}

\title{Quasinormal modes of Charged Regular  Black Hole}

\author{L. A. L\'opez}
\email{lalopez@uaeh.edu.mx}
\author{Valeria  Hinojosa}
\email{hinojosa@inaoep.mx}
\affiliation{\'Area Acad\'emica de Matem\'aticas y F\'isica, UAEH, \\
Carretera Pachuca-Tulancingo Km. 4.5, C. P. 42184, Pachuca, M\'exico.}
\affiliation{Instituto Nacional de Astrof\'isica, \'Optica y Electr\'onica  \\
Departamento de \'Optica, Luis Enrique Erro no1, Tonantzintla, C.P. 72840, Puebla, M\'exico }

\begin{abstract}

The quasinormal modes (QNMs) of a regular black hole with charge are calculated in the eikonal approximation. In the eikonal limit, the QNMs of the black hole are determined by the parameters of the  unstable circular null geodesics. The behavior of the QNMs are compared with the QNMs of Reisner-Nordstr\"{o}m black hole, by fixing some of the parameters that characterize the black holes and varying other. We observed that the parameter that is related to the effective cosmological constant at small distances, determines the behavior of the QNMs of the regular black hole with charge.

\end{abstract}


\maketitle

\section{Introduction}

A problem of the classical theory of gravity is inevitable the existence of singularities. For example, the solutions that describe black holes (BH) as Schwarzschild, Reisner-Nordstr\"{o}m and Kerr have curvature singularities in their interior, although they are covered by an event horizon however, the prediction of Hawking radiation where the negative energy flux of the radiation of black holes cause that the black hole to shrink until the singularity is reached.

The presence of singularities is generally regarded as indicating the breakdown of the theory, requiring modifications that presumably include quantum theory. It is known that at high densities of matter, quantum effects become important, and the matter pressure may be able to counterbalance gravitational collapse and it seems reasonable that when matter reaches Plack density is the onset of quantum gravity effects and there would be enough pressure as to prevent the formation of a singularity. This situation has motivated the derivation and study of non-singular or regular black holes.

To avoid the black hole singularity problem the construction of regular (i.e., nonsingular) solutions have been proposed, for example, the theory of general relativity coupled to nonlinear electrodynamics is a candidate, because in this theory for example, exist the regular magnetic black hole proposed by  Bardeen \cite{Bardeen1} and the regular solution of Bronnikov \cite{Bronnikov:2000vy}.

Another idea to generate regular solutions is to consider the curvature invariants are uniformly restricted by some value which depends only on the type of the curvature invariant. This assumption, called the limiting curvature conjecture was proposed in \cite{Markov:1984ii} and \cite{Polchinski:1989ae}. In general, a regular solution will contain critical scale, mass and charge parameters.

 Hayward \citep{Hayward:2005gi} proposed a regular space-time that describes the formation of a  black hole from an initial vacuum region, its quiescence as a static region (behaving as a cosmological constant at small radius), and its subsequent evaporation to a vacuum region that is a  Bardeen-like static region.  In this context, variations of the Hayward solution have also been proposed as the rotating Hayward \cite{Amir:2015pja} and  Hayward with charge \cite{Frolov:2016pav}.

The black hole always interacts with matter and fields around and as the result of these interactions, it takes a perturbed state regardless of whether or not it is regular. The behavior of the perturbations is of special interest in the stability under perturbations in the linear regime. Perturbed BH is characterized by gravitational waves with complex frequencies that are called quasinormal modes (QNMs) \citep{Kokkotas:1999bd}. The QNMs resonances are crucial to identify the behavior of the black hole parameters, especially the mass and angular momentum of the BH, but as well are going to be important on identifying arising additional physical parameters.

Mashhoon and Ferrari \cite{Mashhoon:1985cya} \cite{Ferrari:1984zz} have suggested an analytical technique of calculating the QNMs in the geometric-optics (eikonal) limit. The basic idea is to interpret the BH free oscillations in terms of null particles trapped at the unstable circular orbit and slowly leaking out. In this sense, Cardoso \cite{Cardoso:2008bp} showed  the relationship among unstable null geodesics, Lyapunov exponents and quasinormal modes in a stationary spherically symmetric space-time.

In \cite{Fernando:2012yw} applying the ideas of Cardoso,  calculated the QNMs frequencies of the regular magnetic BH model proposed by  Bardeen. Also in \cite{Toshmatov:2015wga} the QNMs of regular black holes are studied using the sixth order WKB approximation.

The organization of this paper is as follows. First, in section II a short summary of the Quasi-normal modes and  Lyapunov exponent is given. In Section III the Hayward black hole with charge is described and the effective potential and the shadow are shown for null geodesics and then in section IV, we analyze the QNMs frequencies, in each case the QNMs frequencies are compared with the QNMs of the Reissner-Nordstr\"{o}m BH. Finally, we conclude in Section V with a brief discussion.

\section{Quasi-normal modes and  Lyapunov exponent }

The connection between the QNMs and bound states of the inverted black hole effective potential was pointed out in \cite{Ferrari:1984zz}. In \cite{Cardoso:2008bp} it is shown that, in the eikonal limit,  the QNMs of black holes are determined by the parameters of the circular null geodesics. The real part of the complex QNMs frequencies is determined by the angular velocity at the unstable null geodesics. The imaginary part is related to the instability time scale of the orbit, and therefore related to the Lyapunov exponent that is its inverse. In the case of stationary, spherically symmetric space-times it turns out that this exponent can be expressed as the second derivative of the effective potential evaluated at the radius of the unstable circular null orbit.
It was also shown the  agreement of the so calculated QNMs with the analytic WKB approximation,

\begin{equation}\label{QNM}
\omega_{QNM}=\Omega_{c}\widehat{l} - i (n+\frac{1}{2})| \lambda|,
\end{equation} 
where $n$ is the overtone number and $\widehat{l}$ is the angular momentum of the perturbation. $\Omega_{c}$ is the angular velocity at the unstable null geodesic and  $\lambda$ is the Lyapunov exponent, determining the instability time scale of the orbit. From the equations of motion for a test particle in the static spherically symmetric  space-time, $\dot{r}^{2} + V_{ef}(r)=0$, where $ V_{ef}(r)$  is the effective potential for radial motion, circular geodesics are determined from the conditions $ V_{ef}(r_{c})= V_{ef}^{'}(r_{c})=0$ where $r_{c}$ is the radius of the circular orbit. 

The Lyapunov exponent in terms of the second derivative of the effective potential is given by,

\begin{equation}\label{expLyapunov}
\lambda=\sqrt{\frac{- V_{ef}^{''}(r)}{2\dot{t}^{2}}},
\end{equation}

where $t$ is the time coordinate. The dot denotes the derivative with respect to an affine parameter and the prime stands for the derivative with respect to $r$.

The orbital angular velocity is given by;

\begin{equation}\label{angular}
\Omega_{c}=\frac{d{\varphi}}{d{t}}=\frac{\dot{\varphi}}{\dot{t}}.
\end{equation}

For our purpose both expressions (\ref{expLyapunov}) and (\ref{angular}) should be evaluated at $r_c$, the radius of the unstable null circular orbit. For a static  spherically symmetric background

\begin{equation}\label{sss}
ds^{2}=-f(r)dt^{2}+\frac{1}{g(r)}dr^{2}+r^{2}d\Omega^{2},
\end{equation}

for equatorial orbits  ($\theta= \pi/2$) the Lagrangian is;
\begin{equation}
\mathcal{L}=-f(r)\dot{t^{2}}+\frac{1}{g(r)}\dot{r^{2}}+r^{2}\dot{\phi^{2}}=\delta _1.
\end{equation}

Here $\delta _1= 1, 0$ for spacelike and null geodesics, respectively; the energy  $E$ $(\dot{t}=\frac{-E}{g_{tt}})$ and the angular momentum $L$ $(\dot{\phi}=\frac{L}{g_{\phi\phi}})$ of a test particle  are conserved quantities then for the case of the static spacetime in (\ref{sss}), the effective potential is  given by;

\begin{equation}\label{veg}
   V_{ef} = g(r)\left[ -\frac{E^2}{f(r)}+\frac{L^2}{r^2}-\delta _1\right]
\end{equation}

If we consider the null geodesics $(\delta _1=  0)$, we obtain;

\begin{equation}\label{segundaderivadanull}
V^{''}_{ef}=\frac{L^{2}g(r_{c})}{r_{c}^{4}f(r_{c})}[r_{c}^{2}f^{''}(r_{c})-2f(r_{c})]
\end{equation}

while the orbital angular velocity, which is proportional to the real part of the QNM frequencies, is given by

\begin{equation}
 \Omega=\sqrt{\frac{f(r_{c})}{2r_{c}^{2}}}.
 \end{equation}

\section{Hayward black hole with charge}

Using the idea of the limiting curvature condition and minimal model, the uncharged Hayward BH was proposed considering that the Einstein tensor  $G_{\mu\nu}=R_{\mu\nu}-\frac{1}{4}g_{\mu\nu}R$ has the cosmological constant from  $G \sim - \Lambda g$ as $r \rightarrow 0$ and $\Lambda =3/l^{2}$ where $l$ (Hubble length) is a convenient encoding of the central energy density, the effect of $l$ is that a repulsive force (repulsive core) prevents the singularity. A consequence of including the repulsive core is that the strong energy condition might be violated see \cite{Perez-Roman:2018hfy}. 

Also Hayward considered a spherically symmetric metric as in (\ref{sss}) with $f(r)=g(r)$. For an asymptotically flat space-time with total mass $m$;

\begin{equation}\label{f1}
f(r) \sim 1+\frac{2m}{r} \;\;\ as\;\;\ r \rightarrow \infty,
\end{equation}

and flatness at the center;

\begin{equation}\label{f2}
f(r) \sim 1-\frac{r^{2}}{l^{2}} \;\;\ as\;\;\ r \rightarrow 0
\end{equation}

On the other hand the Hayward with charge was also proposed considering the limiting curvature condition, in this case they were defined  the quadratic curvature invariants of the Einstein ($G^{2}=G_{\mu\nu}G^{\mu\nu}$) and Weyl ($C^{2}=C_{\mu\nu\alpha\beta}C^{\mu\nu\alpha\beta}$) tensors. The assumptions is that the curvature invariants $|G|$ and $|C|$ are uniformly restricted by some values proportional to $l^{-2}$, where the function metric $f(r)$ behaves as (\ref{f1}) and (\ref{f2}).

\begin{figure}[h]
\begin{center}
\includegraphics [width =0.6 \textwidth ]{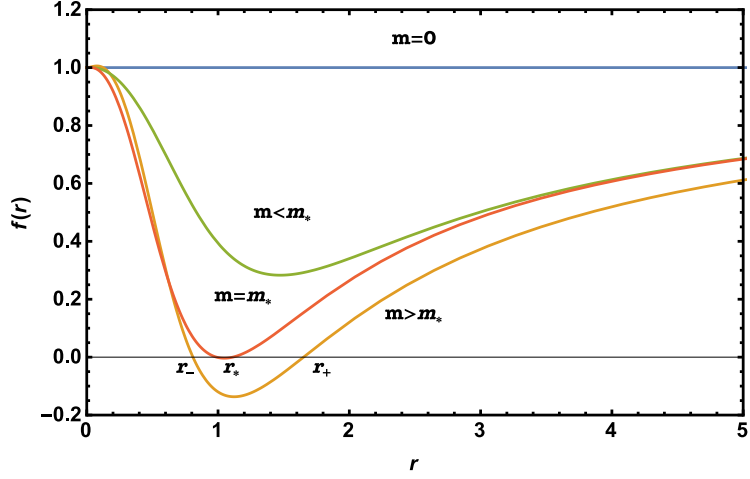}
\end{center}
\caption{Behavior of $f(r)$ for fixed values of parameters $l= 0.5$, $q=1$ and different
values of $m$ with $m_{*}$ the critical mass}\label{Fig1}
\end{figure}

The Hayward black hole with charge (HBH (see \citep{Frolov:2016pav}) is described by the static spherically symmetric space-time (\ref{sss}), where $f(r)=g(r)=\left(1-\frac{(2mr-q^{2})r^{2}}{r^{4}+(2mr+q^{2})l^{2}}\right)$, with $m$ corresponding to mass of the black hole, $q$ is the electric charge and $l$ is a convenient encoding of the central energy density $3/8\pi l^{2} $ assumed positive and  $l$ also is related to an effective cosmological constant at small distances. The metric function of the HBH becomes as Reissner-Nordstr\"{o}m (RN)  black hole when $l\rightarrow 0$. The Hayward black hole with charge is non-singular (regular) at center of the metric where $r\rightarrow 0$; also HBH  is flat for $m =q= 0$. The analysis of $f (r)$ for zeros shows a critical mass $m_{*}=\left(\frac{16(l^{2}+1)^{3})}{27(l^{2}-1)^{4}}q^{6}\right)^{\frac{1}{4}}$. Then if $m_{*}=0$ the spacetime is flat, when $m_{*}=m$ the solution HBH has a horizon (extreme case $r_{*}$);  $m_{*}<m$  has two horizons (the exterior $r_{+}$ and inner $r_{-}$ horizons); finally if   $m_{*}>m$ then $f(r)$ has no zeros. This is shown in Fig. (\ref{Fig1}).

\subsection{Effective Potential and Shadow of HBH}

The corresponding effective potential of null geodesics for the HBH can be calculated from equation (\ref{veg});

 \begin{equation}\label{vefhq}
 V_{ef}=\left(1 - \frac{(2mr - q^{2})r^{2}}{(r^{4}+(2mr + q^{2})l^{2}}\right )\left( \frac{L^{2}}{r^{2}} - \frac{ E^{2}}{\left(1 - \frac{(2mr - q^{2})r^{2}}{r^{4}+(2mr + q^{2})l^{2}}\right)} \right)
 \end{equation}

\begin{figure}[h]
\begin{center}
\includegraphics [width =0.6 \textwidth ]{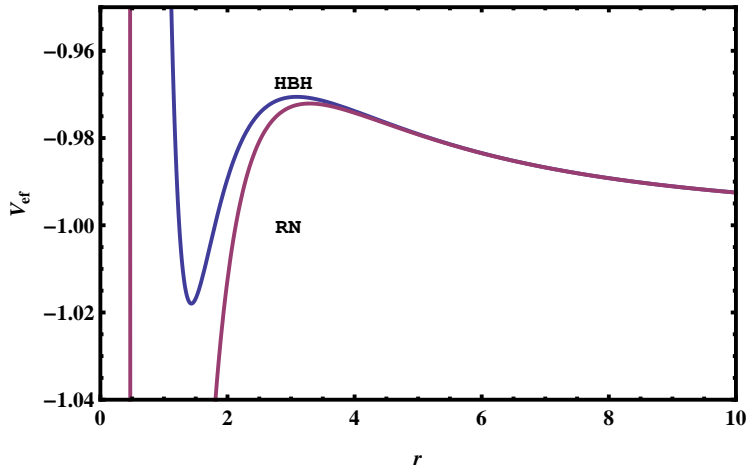}
\end{center}
\caption{In this figure the effective potential as felt by a test particle in Reissner Nordstr\"{o}m (RN), in Hayward black hole with charge (HBH)  and in uncharged Hayward BH (HBH $(q=0)$) are compared.The constants are $m =1.3$, $q =1$, $L =1$, $E =1$ and  $l =0.5$}\label{Vef1}
\end{figure}

When the effective potential is analyzed, it is observed that the effective potential for the HBH is greater than the effective potential of uncharged Hayward BH as it is shown in Fig (\ref{Vef1}).

In Fig (\ref{Vef1}) the comparison between the HBH and Reissner Nordstr\"{o}m effective potentials are shown for null geodesics. The presence of maximum and minimum in the effective potential indicates that there exist circular orbits, stable and unstable for both black holes then the method of Cardoso \cite{Cardoso:2008bp} can be easily applied. Figure (\ref{Vef1}) shows that the effective potential of the Hayward black hole is the greatest compared to Reissner-Nordstr\"{o}m black hole that has a lower potential barrier. Also is possible to observe that there is a difference between  the radius of the circular orbits $r_{c}$ of the black holes.

Also in Fig (\ref{Vef1}) the effective potential of uncharged Hayward BH is shown, as well as in the case of  Hayward black hole with charge exist circular orbits stable and unstable, but the value of the maximum of the potential is lower. Then it is possible to mention that when is introducing the charge to the Hayward black hole the effect of the gravitational potential increases. 
Finally, the effective potential of RN black hole is the greatest compared to uncharged Hayward BH  that has a lower potential barrier.

Consider the radial geodesic when $L = 0$ . Then, $V_{ef}$ is given by $V_{ef}=-E^{2}$ then this shows that the test particle will behave like a free particle for $E = 0$. In  \cite{Abbas:2014oua} the study of the geodesic structure of a regular Hayward black hole is done.

The shadow of a black hole in the region on the observer's sky that is left dark if the light sources are anywhere in the universe but not between the observer and the black hole, the shadow shape will give important information on the parameters of the black hole.

The shadow structure is determined by the properties of null geodesics, in special of circular orbits unstable satisfy $ V_{ef}(r_{c})=0$ and $ V_{ef}^{'}(r_{c})=0$, and we have;

\begin{equation}\label{1}
2f(r_{c})-r_{c}f^{'}(r_{c})=0
\end{equation}

Equation (\ref{1}) does not have any non-trivial solution for HBH. But from the figure (\ref{Vef1}) where the behavior of the effective potentials are shown we observe the existence of photo spheres, since the HBH is spherically symmetric the shadow will be circularly symmetric and will be a function only of the impact parameter $b = L/E$ \cite{Shaikh:2018lcc}. 
Its impact parameter $b_{c}$ is related to the effective potential by $V_{ef}(r_{c})=1/b_{c}^{2}$ i.e., the energy at infinity of those orbits is equal to the maximum of the effective potential.

So it is possible to conclude that the shadow area $(\sigma=\pi b_{c}^{2})$ for HBH, uncharged Hayward BH and RN they have the following behavior; 

\begin{equation}
\sigma_{HBH(q=0)}>\sigma_{RN}>\sigma_{HBH}
\end{equation}

\section{The QNMs of  Hayward black hole with charge}

In this section, we analyzed the quasinormal modes (QNMs)  of the regular black hole with charge in the case of null geodesics and those are then compared with the  QNMs of RN black hole.

\subsection{The real part of the QNMs frequencies}

The angular velocity for massless particles of the Hayward black hole is given by;

 \begin{equation}\label{AngularV}
\Omega=\frac{\sqrt{1-\frac{(2mr-q^{2})r^{2}}{r^{4}+(2mr+q^{2})l}}}{r}
\end{equation}

\begin{figure}[h]
\begin{center}
\includegraphics [width =0.6 \textwidth ]{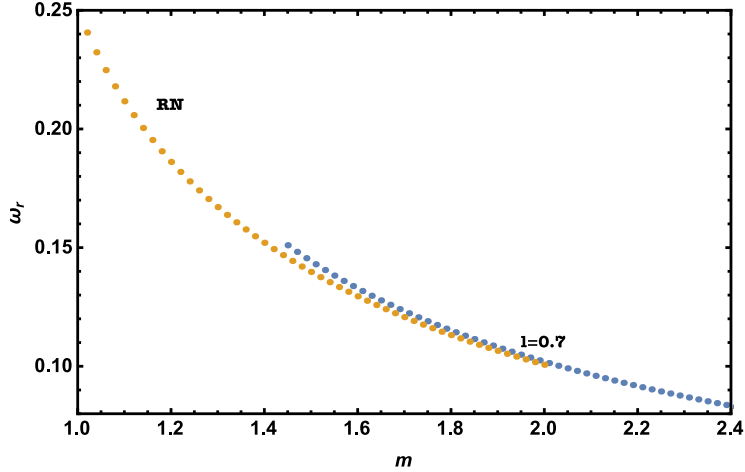}
\end{center}
\caption{QNMs frequencies $\omega_{r}$ of the HBH  and RN are shown as functions of the charge $m$; the other parameters are fixed to $q=1$, $l=0.7$}\label{wrq}
\end{figure}

\begin{figure}[h]
\begin{center}
\includegraphics [width =0.57 \textwidth ]{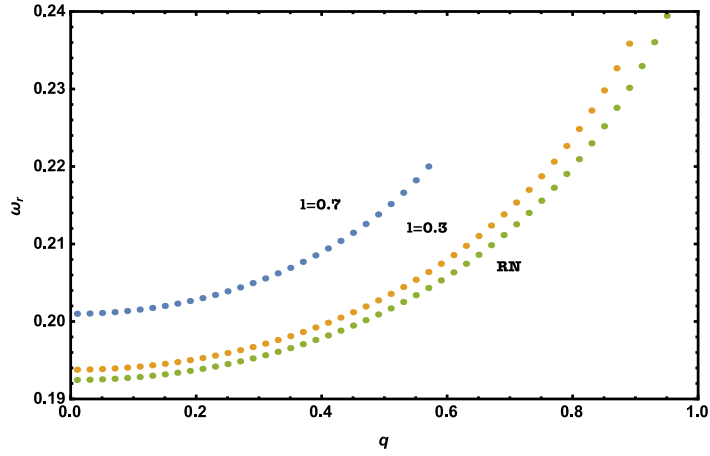}
\end{center}
\caption{QNMs frequencies   $\omega_{r}$ of the HBH  and RN are shown as functions of the charge $q$; the other parameters are fixed to $m=1$, $l=0.7,0.3$}\label{wrm}
\end{figure}

The expression (\ref{AngularV}) should be evaluated at $r_c$, the radius of the unstable null circular orbit.
Hereafter we shall consider $\omega_r/\widehat{l} \mapsto \omega_r$ for our analysis. In Fig. (\ref{wrq}) and (\ref{wrm}) the behavior of the QNMs frequencies  $\omega_{r}$  of the HBH for different values of the parameter $l$  are shown and compared with RN black hole. Both frequencies approach the RN limit as $l$ decreases. When we vary $q$ the frequency $\omega_{r}$ increases when $q$ increases (see Fig (\ref{wrm})). In the other case when we fixed $q$ the QNMs frequencies $\omega_{r}$ decrease when the mass $m$ increases (see Fig (\ref{wrq})).

The period for the circular orbits includes the length of time it takes to the particle to bypass the circular orbit.  For the coordinate time $T_{t}$  is proportional to $1/\omega_{r}$ then when $q$ is fixed the $T_{t}$ increases as $m$ increases,  the opposite occurs when  we vary $q$.

In nonlinear electrodynamics the null geodesics are modified by an effective metric, then the QNMs also are modified as can be consulted in \cite{Breton:2016mqh}. In the table (\ref{Twr}) is illustrated the discrepancy of $\omega_{r}$ between two regular black holes in nonlinear electrodynamics theory and Hayward with charge. The Bardeen black hole (BBH) can be interpreted as the solution to a nonlinear magnetic monopole with mass and charge and the Bronnikov (BrBH) black hole is a regular solution with a nonzero magnetic charge. Nevertheless, the behavior is similar:  $\omega_{r}$ increases when the charge $q$ is increased.

\begin{table}[!hbt]
\begin{center}
\begin{tabular}{|c|c|c|c|}
\hline
 Charge &Hayward BH & Bardeen BH & Bronnikov BH \\
\hline
0.1 & 0.402871 &  0.47158 & 0.385544 \\
\hline
0.2 & 0.405635 &  0.47304  & 0.387518 \\
\hline
0.3 & 0.410495 &  0.472984 & 0.390938 \\
\hline
0.4 & 0.41793 &  0.47422  & 0.39603\\
\hline
0.5& 0.428936 &  0.475814  & 0.403162\\
\hline
0.6 & 0.446173  &  0.477752  & -\\
\hline
\end{tabular}
\caption{The QNMs,  $\omega_{r}$, for the Hayward BH with $\widehat{l}=2$ , $m=1$ and $l=0.7$ are compared with the ones of the effective metric of two regular black holes in nonlinear electrodynamics.}\label{Twr}
\end{center}
\end{table}

When we analyze of the table (\ref{Twr}) data we notice that the values of $\omega_{r}$ for Bardeen are greater in all moment compared with the $\omega_{r}$ of HBH and  Bronnikov BH  ($\omega_{r}(BBH)>\omega_{r}(HBH)>\omega_{r}(BrBH)$).  The period for the circular orbits of HBH is greater than in the case of Bardeen black hole.

\subsection{The imaginary  part of the QNMs frequencies}

We calculate the imaginary part of QNMs frequencies of HBH solution corresponding to different values of parameter $l$ and a comparison with RN black hole is established. Considering the equation (\ref{segundaderivadanull}) we obtained;

\begin{equation}
\label{segdvef}
V^{''}_{ef}=-\left[\frac{2L^{2}(l^{4}(q^{2}+2mr)^{2}+r^{6}(2q^{2}+r(-3m+r))+2l^{2}r^{3}(q^{2}r+m(q^{2}+2r^{2})))}{r^{3}(r^{4}+l^{2}(q^{2}+2mr))^{2}}\right]_{r_{c}}
\end{equation}

Then in the eikonal or geometric-optics limit, the QNMs frequencies $\omega_{i}$ are given by (\ref{QNM}) and (\ref{expLyapunov}).

\begin{figure}[h]
\begin{center}
\includegraphics [width =0.57 \textwidth ]{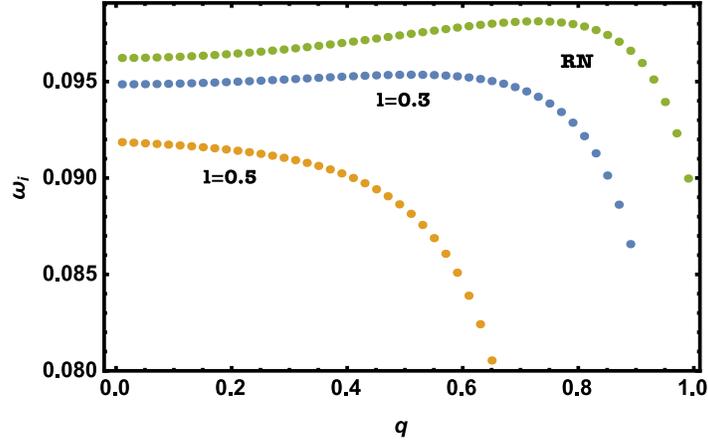}
\end{center}
\caption{QNM frequencies   $\omega_{i}$ of the HBH  and RN are shown as functions of the charge $q$; the other parameters are fixed to $m=1$, $l=0.7,0.3$}\label{wim}
\end{figure}

Now we vary the charge $q$.  The QNMs frequencies $\omega_{i}$ (see Fig (\ref{wim})),  for RN case, increase as $q$ increase and present a maximum and then decrease; for RN the value of $q$ cannot exceed $q=1$ that corresponds to the extreme BH, $q=M$.  HBH does not have this behavior for large values of $l$ and the behavior of  $\omega_{i}$ for HBH is similar to RN for small values of $l$.

\begin{figure}[h]
\begin{center}
\includegraphics [width =0.57 \textwidth ]{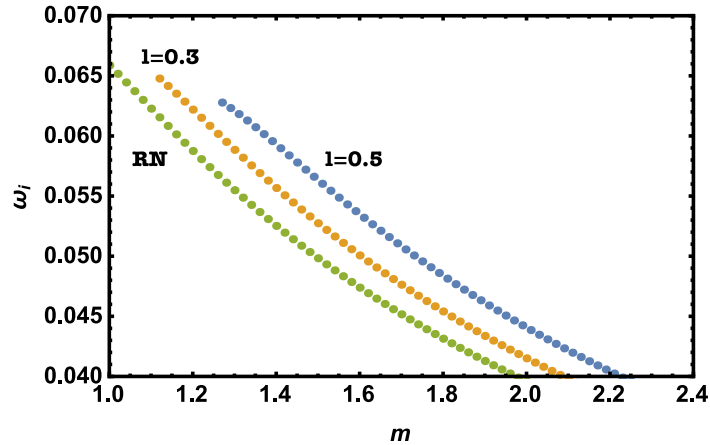}
\end{center}
\caption{QNM frequencies   $\omega_{i}$ of the HBH  and RN are shown as functions of  $m$; the other parameters are fixed to $q=1$, $l=0.5,0.3$}\label{wiq}
\end{figure}

When the mass $m$ varies, $\omega_{i}$ decreases when the mass $m$ increases as shown in Fig (\ref{wiq}) in both BHs.

The instability time scale of the circular null geodesics of HBH  is the greatest compared to Reissner-Nordstr\"{o}m black hole regardless of the value of the mass (see Fig (\ref{wiq})). The opposite occurs when varying the charge $q$ the instability time scale is suppressed as compared with the one of  RN black hole (see Fig (\ref{wim})). In both cases, the instability time tends to zero.

Table \ref{Twi} shows some comparative values to get insight on the discrepancy between the respective frequencies $\omega_{i}$ for two regular black holes in nonlinear electromagnetic and Hayward with charge. The behavior is similar for Bardeen BH and HBH ( when $l$ is increased), in both cases $\omega_{i}$ decreases when the charge $q$  increases. This behavior is in contrast with RN  and Bronnikov BHs; in the case of Bardeen BH  one possible explanation may be the nature of the solution because the charge and mass parameters are not independent, and in fact when the charge is turned off, so does the mass whose origin is purely electromagnetic (see \cite{Breton:2016mqh}), but in the case of Hayward with charge, the parameter $l$ that is a convenient encoding of the central energy density is the cause of such behavior because when $l\rightarrow 0$  the HBH becomes the Reissner-Nordstr\"{o}m (RN)  BH.

\begin{table}[!hbt]
\begin{center}
\begin{tabular}{|c|c|c|c|}
\hline
 Charge &Hayward BH & Bardeen BH & Bronnikov BH \\
\hline
0.1 & 0.091711 &  0.0960758 & 0.0962784 \\
\hline
0.2 & 0.09145 &  0.0956186  & 0.0964377 \\
\hline
0.3 & 0.0909885 &  0.0948243 & 0.0966995 \\
\hline
0.4 & 0.0901247&  0.0936395  & 0.0970555\\
\hline
0.5& 0.088398 &  0.091981  & 0.0974872\\
\hline
0.6 & 0.0845255  &  0.091981  & -\\
\hline
\end{tabular}
\caption{The $\omega_{i}$ for the Hayward black hole with $n=0$, $m=1$ and $l=0.5$ are compared with the $\omega_{i}$ of the effective metric of two regular black holes in nonlinear electromagnetic }\label{Twi}
\end{center}
\end{table}

Also of the Table \ref{Twi}, is possible to observe that the $\omega_{i}$ of the Bronnokov are greater than the  $\omega_{i}$ of HBH in the other words $\omega_{i}(BrBH)>\omega_{i}(BBH)>\omega_{i}(HBH)$ but the HBH instability time scale is suppressed as compared with Bardeen. The HBH is more stable compared to two regular black holes in nonlinear electromagnetic.

\section{Conclusions}

We have studied the QNMs frequencies of the Hayward black hole with charged through the Lyapunov exponent in the optical approximation. QNMs frequencies were calculated from the unstable null geodesics when the charge or mass is varied and in all cases, the comparison is done with the QNMs frequencies of the RN black hole.

It is observed that the effective potential for the HBH is greater than the effective potential of uncharged Hayward BH and the shadow of uncharged Hayward BH is greater than the shadow of the HBH.

When we keep fixed the charge the imaginary and real part of the QNMs decrease when the mass increases; in both BHs the imaginary frequency approaches zero. In the case of fixing the mass the imaginary part of QNMs frequencies for HBH, increases as the charge augment and presents a maximum; the decrease is more notorious for small values of $l$  and approaches the RN limit. The real part of the QNMs increase when the charge increases but the variation of the charge is restricted by the values of the mass and the parameter $l$.

The real $w_{r}$ for HBH is greater than for RN black hole in the case of varying the charge.
However, the interval of variation of $q$ is shorter than the one for RN BH, and is shorter for larger $l$; for $l=0$ the $q$ ranges are the same for HBH and  RN BHs. In the case of the imaginary part, the values of $w_{i}$ for HBH decrease for larger $l$ as compared with the ones of RN black hole. The effective potential of Hayward black hole is the greatest compared to Reissner-Nordstr\"{o}m black hole.

Finally, the analysis of the QNMs shows that for the charged  regular black hole the stability of the test field is modified as well as the period for the circular orbits.  The imaginary frequencies are shorter for the HBH than for RN, implying that the test field is amortiguated longer than in the RN geometry. The real frequencies are larger for HBH than for RN, showing then that the oscillations of the test field are faster than the ones of RN BH. Also are shown some comparative values to get insight on the discrepancy between the respective frequencies of two regular black holes in nonlinear electrodynamics and the HBH with charge.

\bibliographystyle{unsrt}
\bibliography{bibliografia}

\end{document}